\documentclass[conference]{IEEEtran}
\IEEEoverridecommandlockouts
\usepackage[utf8]{inputenc}
\usepackage{cite}
\usepackage{amsmath,amssymb,amsfonts}
\usepackage{algorithmic}
\usepackage{algorithm}
\usepackage{graphicx}
\usepackage{textcomp}
\usepackage{todonotes}
\usepackage{caption}
\usepackage{subfig}
\usepackage{setspace}
\usepackage{times}
\usepackage{hyperref}
\usepackage{todonotes}
\def\BibTeX{{\rm B\kern-.05em{\sc i\kern-.025em b}\kern-.08em
    T\kern-.1667em\lower.7ex\hbox{E}\kern-.125emX}}

\begin{document}

%\title{Arriving On Time: Path-Based Stochastic Routing\\
%\title{Arriving On Time: Stochastic Routing in Path-Centric Uncertain Road Networks\\
%\title{Shortest Path with On-Time Arrival Reliability in Path-Centric Uncertain Road Networks\\
\title{A New Formulation of The Shortest Path Problem with On-Time Arrival Reliability\\% in Path-Centric Uncertain Road Networks\\
%\title{Stochastic Routing in Path-Centric Uncertain Road Networks
%{\footnotesize \textsuperscript{*}Note: Sub-titles are not captured in Xplore and
%should not be used}
}

\author{
%\IEEEauthorblockN{1\textsuperscript{st} Georgi Andonov}
\IEEEauthorblockN{Georgi Andonov}
\IEEEauthorblockA{\textit{Department of Computer Science} \\
\textit{Aalborg University, Denmark}\\
%Aalborg, Denmark \\
gandon16@student.aau.dk}
\and
%\IEEEauthorblockN{2\textsuperscript{nd} Given Name Surname}
\IEEEauthorblockN{Bin Yang}
\IEEEauthorblockA{\textit{Department of Computer Science} \\
\textit{Aalborg University, Denmark}\\
%Aalborg, Denmark \\
byang@cs.aau.dk}
%\and
%\IEEEauthorblockN{Samitha Samaranayake }
%\IEEEauthorblockA{\textit{School of Civil and Environmental Engineering } \\
%\textit{Cornell University}\\
%Ithaca, NY \\
%samitha@cornell.edu}
%\and
}

% \author{
% %\IEEEauthorblockN{1\textsuperscript{st} Georgi Andonov}
% %\IEEEauthorblockN{Georgi Andonov}
% %\IEEEauthorblockA{\textit{Department of Computer Science} \\
% Georgi Andonov and Bin Yang \\
% Aalborg University, Aalborg, Denmark\\
% gandon16@student.aau.dk byang@cs.aau.dk
% }

\maketitle
\thispagestyle{plain}
\pagestyle{plain}
\pagenumbering{arabic}

%\vspace{-10pt}

\begin{abstract}
We study stochastic routing in the PAth-CEntric (PACE) uncertain road network model. In the PACE model, uncertain travel times are associated with not only edges but also some paths.
The uncertain travel times associated with paths are able to well capture the travel time dependency among different edges. This significantly improves the accuracy of travel time distribution estimations for arbitrary paths, which is a fundamental functionality in stochastic routing, compared to classic uncertain road network models where uncertain travel times are associated with only edges.

We investigate a new formulation of the shortest path with on-time arrival reliability (SPOTAR) problem under the PACE model. Given a source, a destination, and a travel time budget, the SPOTAR problem aims at finding a path that maximizes the on-time arrival probability. We develop a generic algorithm with different speed-up strategies to solve the SPOTAR problem under the PACE model. Empirical studies with substantial GPS trajectory data offer insight into the design properties of the proposed algorithm and confirm that the algorithm is effective.

This report extends the paper ``Stochastic Shortest Path Finding in Path-Centric Uncertain Road Networks'', to appear in IEEE MDM 2018, by providing a concrete running example of the studied SPOTAR problem in the PACE model and additional statistics of the used GPS trajectories in the experiments.

%This document investigates the shortest path with on-time arrival reliability (SPOTAR) problem. SPOTAR problem tries to find an a priori path that maximizes the on time arrival reliability at reaching a destination starting from a source within a predefined time budget. We present an algorithm that solves the SPOTAR problem considering the travel costs dependences along the SPOTAR path. Experimental results evaluate the quality of the proposed solution and summarize some limitations of the approach considered in this paper. The data used in the experiments have been gathered by GPS observations from Aalborg.
\end{abstract}

\begin{spacing}{1.5}
\end{spacing}

\section{Introduction}
%The Point-To-Point shortest path is the problem of finding the shortest path between source and destination vertices in a graph. The problem can be solved only if the graph does not contain any negative cycles. One way to solve this problem is by using Dijkstra’s algorithm. Dijkstra’s algorithm has been widely used to find the shortest path between two vertices in a graph. The algorithm runs in super-linear time. Unfortunately, in order to use Dijkstra's algorithm, the edges in a graph must be labeled with deterministic values which is not allays possible.

%If the edges in a graph are labeled with probability functions then the problem is referred to the stochastic routing problem domain.
%Two problems in the area of stochastic routing have been identified in \cite{a3}.

Emerging transportation innovations, such as mobility-on-demand services and autonomous driving, call for stochastic routing where travel time distributions, but not just average travel times, are utilized~\cite{a0, a1}. For example, consider two paths $P_1$ and $P_2$ where both paths connect the same source and destination. The travel time distribution of the two paths are shown in Table~\ref{tbl:paths}.
\begin{table}[!htp]
	\centering
	\begin{tabular}{|c|c|c|c|c|}
		\hline
		Travel time (mins) & 40 & 50 & 60 & 70 \\ \hline\hline
		$P_1$ & 0.5 & 0.2 & 0.2 & 0.1 \\ \hline
		$P_2$ & 0 & 0.8 & 0.2 & 0 \\\hline
	\end{tabular}
	\caption{Travel Time Distributions for $P_1$ and $P_2$}
	\label{tbl:paths}
\end{table}

If only considering average travel times, $P_1$ is always the best choice, because $P_1$ has average travel time 49 whereas $P_2$'s average travel time is 52. However, if a person needs to go to the destination within 60 minutes, e.g., to catch a flight, then $P_2$ is the best choice as it guarantees that the person will arrive the destination within 60 minutes. In contrast, taking $P_1$ may run into a risk that the person will miss the flight when the path takes 70 minutes.

In this paper, we consider \emph{stochastic routing} that takes into account travel time distributions. In particular, we investigate the shortest path with on-time arrival reliability (SPOTAR) problem. Given a source, a destination, and a travel time budget, e.g., 60 minutes, the SPOTAR problem aims at finding a path that maximizes the on-time arrival probability, i.e., the probability of arriving the destination within the time budget.

Although the SPOTAR problem has been studied in the literature~\cite{a2, a3}, they all consider a classical uncertain road network modeling, where uncertain travel times are assigned to only edges, and the uncertain travel times on different edges are independent. We call the classical model the \emph{edge-centric} model.
However, recent studies suggest that the travel times on different edges are often highly dependent~\cite{a1, a1-1}. The edge-centric model ignores the dependency and thus results in inaccurate travel time distribution computation for paths, especially for long paths. To contend with this, a PAth-CEntric model (PACE) has been proposed~\cite{a1, a1-1}. In the PACE model, not only edges, but some paths are also associated with uncertain travel times and the uncertain travel times that are associated with paths well capture the dependency of the travel times among different edges in the path. Thus, the PACE model is able to provide much more accurate travel time distributions~\cite{a1, a1-1}. However, existing algorithms that work for the edge-centric model cannot be applied directly in the new PACE model. In this paper, we investigate how to solve the SPOTAR problem under the PACE model.

\textbf{Contributions: }To the best of our knowledge, this is the first paper to study the SPOTAR problem in the PACE model that
%This article propose a innovative way of solving the SPOTAR problem by
exploits travel time dependencies among edges.
First, we define the SPOTAR problem in the PACE model.
Second, we propose a generic algorithm with different speed-up heuristics to solve the problem.
Third, we report on comprehensive experiments based on real world trajectory data.

\section{Related work}
We first review two uncertain road network models and then discuss existing studies on stochastic shortest path finding.

\textbf{Uncertain Road Network Modeling: }The edge-centric uncertain road network model has been applied extensively in stochastic routing~\cite{a2,a3,a5,a6,a7,a71,a72,a73,a74,a75,a76,a8}. In the edge-centric model, each edge is associated with an uncertain travel time and different edges' uncertain travel times are independent.
Given a path, the uncertain time of the path is derived by convoluting the uncertain times of the edges in the path.
%
%for solving problems in the stochastic routing field applies incrementally convolution of probability distributions to all edges of a path starting from the source towered the destination. This model has been studied in \cite{a2,a3,a5,a6,a7,a8}, unfortunately, the model lacks accuracy because it does not take into account the dependencies between the segments in a path.
%
However, uncertain travel times among different edges in a path are often dependent but not independent. The edge-centric model is unable to capture the dependency and thus results in inaccurate uncertain travel times for paths.
To contend with this, the PAth-CEntric (PACE) model is proposed recently~\cite{a1, a1-1}. PACE also associates some paths with joint travel time distributions that well capture the dependency of the uncertain travel times of the edges in the paths.
%
%in relation to Stochastic routing has been studied in.
%The articles suggest the usage of trajectory data to derive travel cost from joint distribution of a path by taking into account the travel cost dependencies.

\textbf{Stochastic shortest path finding: } There exist two categories of stochastic shortest path finding---the Shortest path with on-time arrival reliability problem (SPOTAR) and the Stochastic on time arrival problem (SOTA)~\cite{a2,a3}.
%
%can find the best arc to follow at each junction,
%
SPOTAR identifies an a-priori path that maximizes the on time arrival reliability of reaching a destination within a pre-defined time budget.
SOTA considers a scenario where a vehicle keeps moving and needs to choose which edge to follow at each intersection, based on the travel time has been already spent. SOTA tries to find an optimal policy that guides a vehicle to choose the next optimal edge at each intersection. %, to choose an edge to follow based on the travel time has been already spent.
%In this sense, SPOTAR considers a static setting where SOTA considers a dynamic setting.
We study the SPOTAR problem in the PACE model, while exsiting studies on SPOTAR all consider the edge-centric model.
%
%SOTA problem is usually defined as a dynamic programing problem and then solved as described by \cite{a4,a5,a6}.

%This article focuses on the SPOTAR problem because our goal is to identify an a priori path that results in the highest probability of reaching a destination node from a source node within a given time budget in an uncertain road network.

%A general solution of the SPOTAR problem has been given by Nie and Wu \cite{a4}, the solution uses a first-order stochastic dominance of paths. In the worst case, the approach runs in exponential time. However, Nie and Wu assumes that the travel costs of links in the network are independent. This is in contrast with our proposal which aims to exploit the travel cost dependencies in a path.

% \begin{spacing}{1.0}
% \end{spacing}

%\section{Problem formulation}
\section{Preliminaries}

\begin{figure}[!htp]
  \begin{center}
\includegraphics[width=85mm,scale=0.5]{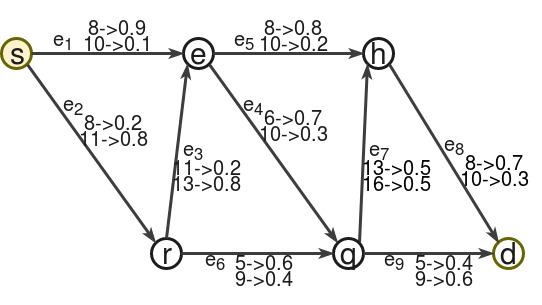}
  \end{center}
   \caption{Road Network and Uncertain Edge Weights}
	\label{fig:origin_graph}
\end{figure}

We use a graph to model a road network. In particular, a road network  is modeled as a directed graph $G=(V, E)$. $V$ is a set of nodes that represent intersections and $E \subseteq V \times V$ is a set of edges that represent road segments.
For example, Figure~\ref{fig:origin_graph} shows a graph that models a road network with 6 intersections (i.e., nodes) and 9 road segments (i.e., edges).

A path is a sequence of adjacent edges $P = \langle e_1, e_2$, $\ldots$, $e_n\rangle$, where $n \geq 1$.
In addition, we require that the edges in path $P$ are unique, meaning that $e_i\neq e_j$, if $1 \leq i, j \leq n$ and $i\neq j$.
%
%Let the set $PATH_P$ consists of all arcs of a given path p or more formally $\forall a_i \in PATH_P$ for $i \in [ 1,n )$, we have that $PATH_P \subseteq A $. Let $v_i$ be the start vertex of arc $a_i$, let $v_{i+1}$ be the destination vertex of $a_i$, therefore $\forall a_k, a_{k+1} \in PATH_P, k \in [ 1,n -1 )$ there is a vertex $v_{k+1}$. The number of elements in the set $PATH_P$ which by definition is $n = |PATH_P|$ will be referred as a cardinality of a path p.
%
Next, we define a subsequence of edges in path $P$ as its sub-path $P'$. Formally, $P' = \langle e_i, e_{i+1}$, $\ldots$, $e_{j-1}$, $e_j\rangle$, where $1 \leq i<j\leq n$.
%  consists of sequence of arcs which are all elements of the set $PATH_P$. Let $PATH_{PS}$ be the set of all arcs for a given sub-path $p_{sub}$. Defined formally as follows $\forall a_{k} \in PATH_{PS} ,  k \in [ i,j ]$ , we have that $PATH_{PS} \subseteq PATH_{P} \subseteq A$.
%
For example, in Figure~\ref{fig:origin_graph}, path $P_1=\langle e_1, e_4 \rangle$ is a sub-path of path $P_2=\langle e_1, e_4, e_9 \rangle$.

Based on the above definitions, we proceed to introduce the classic edge-centric uncertain road network model and the path-centric uncertain road network model (i.e., PACE), respectively.

\emph{Edge-centric uncertain road network model: } We maintain a weight function $W$ that takes as input an edge and returns an uncertain travel time for the edge. We maintain an uncertain travel time for each edge. The edge-centric model is considered as the de-facto model in stochastic routing~\cite{a2, a3, a4, a5, a6, a7, a8}. %This means that the classical model only maintains edge weights.
%%In this paper, we use histograms to represent uncertain travel times.

The uncertain edge weights are often instantiated using trajectories that occurred on the road network. Given a set of trajectories, we first break the trajectories to small pieces that fit the underlying edges and then use the small pieces to instantiate edge weights.

Assume that there we have 100 trajectories as shown in Table~\ref{tbl:traj}.
The first row of Table~\ref{tbl:traj} means that 80 trajectories traversed path $\langle e_1, e_4 \rangle$, which spent 8 mins on $e_1$ and 6 mins on $e_4$.
%
% \begin{table}[!htp]
% \centering
% %\scriptsize
% \begin{tabular}{|c|l|c|} \hline
% Traversed Path & Costs on edges & \# of trajectories \\ \hline \hline
% $\langle e_1, e_4 \rangle$ & 8, 6 & 72 \\ \hline
% $\langle e_1, e_4 \rangle$ & 8, 10 & 18 \\ \hline
% $\langle e_1, e_4 \rangle$ & 10, 6 & 8 \\ \hline
% $\langle e_1, e_4 \rangle$ & 10, 10 & 2 \\ \hline
% $\langle e_4, e_9 \rangle$ & 6, 5 & 60 \\ \hline
% $\langle e_4, e_7 \rangle$ & 10, 13 & 40 \\ \hline
% \end{tabular}
% \caption{Trajectory Examples}\label{tbl:traj}
% \end{table}
%
\begin{table}[!htp]
\centering
%\scriptsize
\begin{tabular}{|c|l|c|} \hline
Traversed Path & Costs on edges & \# of trajectories \\ \hline \hline
$\langle e_1, e_4 \rangle$ & 8, 6 & 80 \\ \hline
$\langle e_1, e_4 \rangle$ & 10, 10 & 20 \\ \hline
$\langle e_1 \rangle$ & 8 & 100 \\ \hline
\end{tabular}
\caption{Trajectory Examples}\label{tbl:traj}
\end{table}

Now, we are able to instantiate $W(e_1)=\{(8, 0.9)$, $(10, 0.1)\}$, meaning that $e_1$ may take 8 mins with probability 0.9, and 10 mins with probability 0.1, as shown in Fig.~\ref{fig:origin_graph}.
This is because out of 200 trajectories that traversed $e_1$, 180 trajectories took 8 mins and 20 trajectories took 10 mins. Similarly, we have $W(e_4)=\{(6, 0.8), (10, 0.2)\}$.

% Bin: old text using the old trajectory table
% Based on these trajcectories, we are able to instantiate $W(e_1)=\{(8, 0.9), (10, 0.1)\}$, meaning that $e_1$ may take 8 mins with probability 0.9, and 10 mins with probability 0.1, as shown in Fig.~\ref{fig:origin_graph}. This is becasue out of 100 trajectories that traversed $e_1$, 90 trajectories took 8 mins and 10 trajectories took 10 mins. Similarly, we have $W(e_4)=\{(8, 0.9), (10, 0.1)\}$, becasue out of 200 trajectoris that traversed $e_4$, 140 trajectories took 6 mins and the remaining 60 took 10 mins.

In this edge-centric model, the uncertain travel times of edges are independent. Thus, the uncertain travel time of a path is computed as the convolution of the uncertain travel times of the edges in the path~\cite{a2, a3, a4, a5, a6, a7, a8}.
For example, assuming that we have instantiated the edge weights as shown in Figure~\ref{fig:origin_graph}.
Consider path $P_1=\langle e_1, e_4\rangle$. Since we assume that the travel times of $e_1$ and $e_4$ are independent, we compute the joint travel time distribution first. Then, based on the joint distribution, we are able to compute the total travel time distribution, which is shown in Table~\ref{tbl:dist}.
\begin{table}[!htp]
	\centering
	\subfloat[Joint Travel Time Distribution]{
		\begin{minipage}[c]{0.24\textwidth}
			\begin{tabular}{|c|c|c|c|} \hline
             $e_1$ & $e_4$ & Probability \\ \hline \hline
             8 & 6 & 0.63 \\ \hline
             8 & 10 & 0.27 \\ \hline
             10 & 6 & 0.07 \\ \hline
             10 & 10 & 0.03 \\ \hline
            \end{tabular}
		\end{minipage}
	}
	\subfloat[Total Travel Time Distribution]{
		\begin{minipage}[c]{0.24\textwidth}
			\begin{tabular}{|c|c|} \hline
             $P_1$ & Probability \\ \hline \hline
             14 & 0.63 \\ \hline
             18 & 0.27 \\ \hline
             16 & 0.07 \\ \hline
             20 & 0.03 \\ \hline
            \end{tabular}
		\end{minipage}
	}
	\caption{Joint vs. Total Travel Time Distribution}
	\label{tbl:dist}
\end{table}

% Bin: old table
% \begin{table}[!htp]
% 	\centering
% %	\subfigure[Joint Cost Distribution]{
% 		\begin{minipage}[b]{0.24\textwidth}
% 			\begin{tabular}{|c|c|c|c|} \hline
%              $e_1$ & $e_4$ & Probability \\ \hline \hline
%              8 & 6 & 0.63 \\ \hline
%              8 & 10 & 0.27 \\ \hline
%              10 & 6 & 0.07 \\ \hline
%              10 & 10 & 0.03 \\ \hline
%             \end{tabular}
% 		\end{minipage}
% 	%}
% 	%\subfigure[Total Cost Distribution]{
% 		\begin{minipage}[b]{0.24\textwidth}
% 			\begin{tabular}{|c|c|} \hline
%              $P_1$ & Probability \\ \hline \hline
%              14 & 0.63 \\ \hline
%              18 & 0.27 \\ \hline
%              16 & 0.07 \\ \hline
%              20 & 0.03 \\ \hline
%             \end{tabular}
% 		\end{minipage}
% 	%}
% 	\caption{Joint vs. Total Travel Time Distribution}
% 	\label{tbl:dist}
% \end{table}

Note that the computed total travel time distribution of path $P_1$ is inconsistent with the ``ground truth'' distribution that is reflected in the trajectories in Table~\ref{tbl:traj}. The ground truth distribution suggests $P_1$ took 14 mins with probability 0.8 (because out of 100 trajectories, 80 trajectories took 14 mins to traverse $P_1$) and 20 mins with probability 0.2 (because out of 100 trajectories, 20 trajectories took 20 mins to traverse $P_1$).
This example shows that the travel time dependency, i.e., traveling fast/slow on $e_1$ and also traveling fast/slow on $e_2$, is broken when using the edge-centric model, which results in inaccurate uncertain travel time for paths. It also shows that it is very important to use trajectories that are traversing the whole path, but not only intendant edges when calculating the travel cost distribution of a path.

% \begin{table}[!htp]
% \centering
% %\scriptsize
% \begin{tabular}{|c|c|c|c|} \hline
%  $e_1$ & $e_4$ & Probability \\ \hline \hline
%  8 & 6 & 0.63 \\ \hline
%  8 & 10 & 0.27 \\ \hline
%  10 & 6 & 0.07 \\ \hline
%  10 & 10 & 0.03 \\ \hline
% \end{tabular}
% \caption{Joint Cost Distribution of $P_1=\langle e_1, e_4\rangle$}\label{tbl:jointdist}
% \end{table}

% \begin{table}[!htp]
% \centering
% %\scriptsize
% \begin{tabular}{|c|c|} \hline
%  $P_1$ & Probability \\ \hline \hline
%  14 & 0.63 \\ \hline
%  18 & 0.27 \\ \hline
%  16 & 0.07 \\ \hline
%  20 & 0.03 \\ \hline
% \end{tabular}
% \caption{Total Distribution of $P_1=\langle e_1, e_4\rangle$}\label{tbl:jointdist}
% \end{table}

\emph{PAth-CEntric (PACE) uncertain road network model: } In the PACE model~\cite{a1, a1-1}, not only edges are associated with uncertain travel times, some paths are also associated with uncertain travel times. The uncertain travel times that are associated with paths model the joint travel time distributions of the edges in the paths.

Following the running example, PACE maintains not only uncertain travel times for edges, i.e., $W(e_1)$ and $W(e_4)$, but also the joint travel time distribution for path $P_1$, i.e., $W(P_1)$, as shown Table~\ref{tbl:PACEdist}(a). The joint travel time distributions maintained in PACE are directly derived from trajectories.
Based on $W(P_1)$, we are able to derive total travel time distribution of $P_1$, which well aligns with the ground truth distribution.
\begin{table}[!htp]
	\centering
	\subfloat[$W(P_1)$]{
		\begin{minipage}[c]{0.24\textwidth}
			\begin{tabular}{|c|c|c|c|} \hline
             $e_1$ & $e_4$ & Probability \\ \hline \hline
             8 & 6 & 0.8 \\ \hline
             10 & 10 & 0.2 \\ \hline
            \end{tabular}
		\end{minipage}
	}
	\subfloat[Total Travel Time Distribution]{
		\begin{minipage}[c]{0.24\textwidth}
			\begin{tabular}{|c|c|} \hline
             $P_1$ & Probability \\ \hline \hline
             14 & 0.8 \\ \hline
             20 & 0.2 \\ \hline
            \end{tabular}
		\end{minipage}
	}
	\caption{Joint Distributions in PACE}
	\label{tbl:PACEdist}
\end{table}

Note that, in PACE, only some paths are associated with uncertain travel times, but not all paths. A road network may have a huge number of meaningful paths and thus we cannot afford maintain uncertain travel times for all paths in PACE. In addition, we often lack sufficient trajectories to cover all paths. In practice, we maintain joint distributions for those paths which have been traversed by sufficient amount of trajectories, e.g., popular paths.

Next, we present how to compute the travel time distribution of a path in PACE by using a concrete example.
Assuming that, in addition to path travel time distribution $W(P_1)$, PACE also maintains path travel time distribution $W(P_2)$ for path $P_2=\langle e_4, e_9\rangle$. Then,
given a path $P=\langle e_1, e_4, e_9\rangle$, there exist more than one combination of weights such that each combination covers $P$. %a unique set of paths which can cover the path.
For example, we may use $\{W(e_1)$, $W(e_4)$, $W(e_9)\}$, $\{W(e_1)$, $W(P_2)\}$, $\{W(P_1)$, $W(e_9)\}$, and $\{W(P_1)$, $W(P_2)\}$ to compute $P$'s uncertain travel time, respectively.
It has been shown that the coarsest combination, i.e., the combination with the longest sub-paths, gives the most accurate uncertain travel time~\cite{a1, a1-1}. In our example, it is $\{W(P_1)$, $W(P_2)\}$.
%
%An efficient algorithm that identifies the coarest combination from avaiable path and edge weigths is also provided in our previous study.

After identifying the coarsest combination for a query path $P$, say $\{P_1, P_2, \ldots, P_n\}$, we use Equation~\ref{eqn_pace} to compute the joint distribution of $P$~\cite{a1, a1-1}.
\begin{equation}
\label{eqn_pace}
prob(P) = \frac{\Pi_{i=1}^{n}W(P_i)}{\Pi_{i=1}^{n-1}W(P_i\cap P_{i+1})}
\end{equation}
where $P_i\cap P_{i+1}$ denotes the overlapped path of paths $P_i$ and $P_{i+1}$.
Consider the running example where $P=\langle e_1, e_4, e_9\rangle$. We have $prob(P)=\frac{W(P_1)\cdot W(P_2)}{W(\langle e_4 \rangle)}$ since $P_1 \cap P_2=\langle e_4 \rangle$.

\textbf{Problem Definition: } Given a source vertex $v_s$, a destination vertex $v_d$, and a travel time budget $T$, we aim at identifying path $P^*$ which goes from $v_s$ to $v_d$ and has the largest probability of arriving the destination $v_d$ within the time budget $T$. This means that $P^*=\arg \max_{P\in \mathit{Path}} \mathit{Prob}(P.\mathit{TravelTime}\leq T)$, where set $\mathit{Path}$ contains all paths from $v_s$ to $v_d$ and the travel time distribution of path $P$ is computed using the PACE model.

\section{Proposed Solution}

We propose an algorithm for solving the SPOTAR problem based on the PACE model. The algorithm is based on a heuristic function that estimates the least possible travel time from a vertex to the destination vertex.
%
%it uses a heuristic to estimate the least possible travel time that is needed to travel from a vertex to the destination vertex.
We first introduce the heuristic function and then introduce the algorithm using the heuristic function.

\subsection{Heuristic Function}
\label{su:heuristic}
%Our proposal for solution is very similar to A* search. Since the search space is exponentially large, the proposal uses a heuristic to prune the search space and speed up the computation. In this way, we explore only small part of the search space which leads to improved performance compare to the same approach without pruning.

Motivated by $A^*$ algorithm, to decide which edge we need to explore next to find the SPOTAR path, we maintain a heuristic function that takes as input a vertex and returns \emph{the least possible travel time} from the vertex to the destination vertex $v_d$.

A baseline heuristic function can simply return a travel time that equals to the Euclidean distance between the argument vertex and $v_d$ divided by the maximum speed limit in the road network. However, this heuristic is too optimistic and thus loose. We aim at introducing a tighter heuristic.

\begin{figure}[!htp]
  \begin{center}
\includegraphics[width=70mm,scale=0.5]{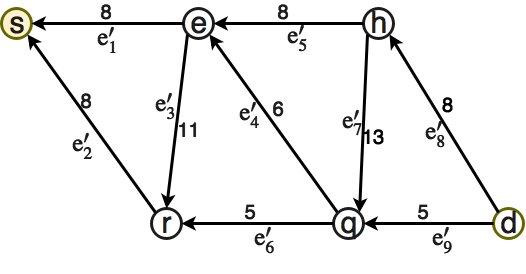}
  \end{center}
   \caption{Reversed Graph $G_{rev}$ of $G$ }
  \label{fig:min_travel_graph}
\end{figure}

To this end, we first introduce a reversed graph $G_{rev}=(V, E')$ based on the original graph $G=(V, E)$ and then compute a shortest path tree that is rooted at $v_d$ in $G_{rev}$.
In particular, $G_{rev}$ contains the same vertices with $G$ and the edges in $G_{rev}.E'$ have reversed directions of the edges in $G.E$. Fig.~\ref{fig:min_travel_graph} shows the reversed graph of the graph shown in Fig.~\ref{fig:origin_graph}. In the reversed graph, the edge weights are deterministic. In particular, for an edge $e'=(v_i, v_j)\in E'$, its weight equals to the least travel time of the uncertain travel time on edge $e=(v_j, v_i) \in E$ in the original graph.
For example, edge $e_1^\prime$ in $G_{rev}$ has weight 8 since the least travel time of edge $e_1$ in $G$ is 8.

Next, we compute a shortest path tree that is rooted as the destination vertex $v_d$ in the reversed graph $G_{rev}$ using Dijkstra's algorithm.
%
%To compute the heuristic, initially a backward Dijkstra search calculates a shortest path tree starting from the destination vertex $d$ in $G_{rev}$ which is the reversed graph of $G$.
%
Note that Dijkstra's algorithm finishes when the distance from the destination vertex to a vertex becomes higher than the time budget $T$.
In other words, we do not need to compute a full shortest path tree but only part of the shortest path tree that contains the vertices whose distances to $v_d$ are not larger than $T$.
We use function $v_i.getMin()$ to return the minimum travel time which is needed to travel from $v_i$ to destination $v_d$.

\begin{figure}[!htp]
  \begin{center}
\includegraphics[width=67mm,scale=0.5]{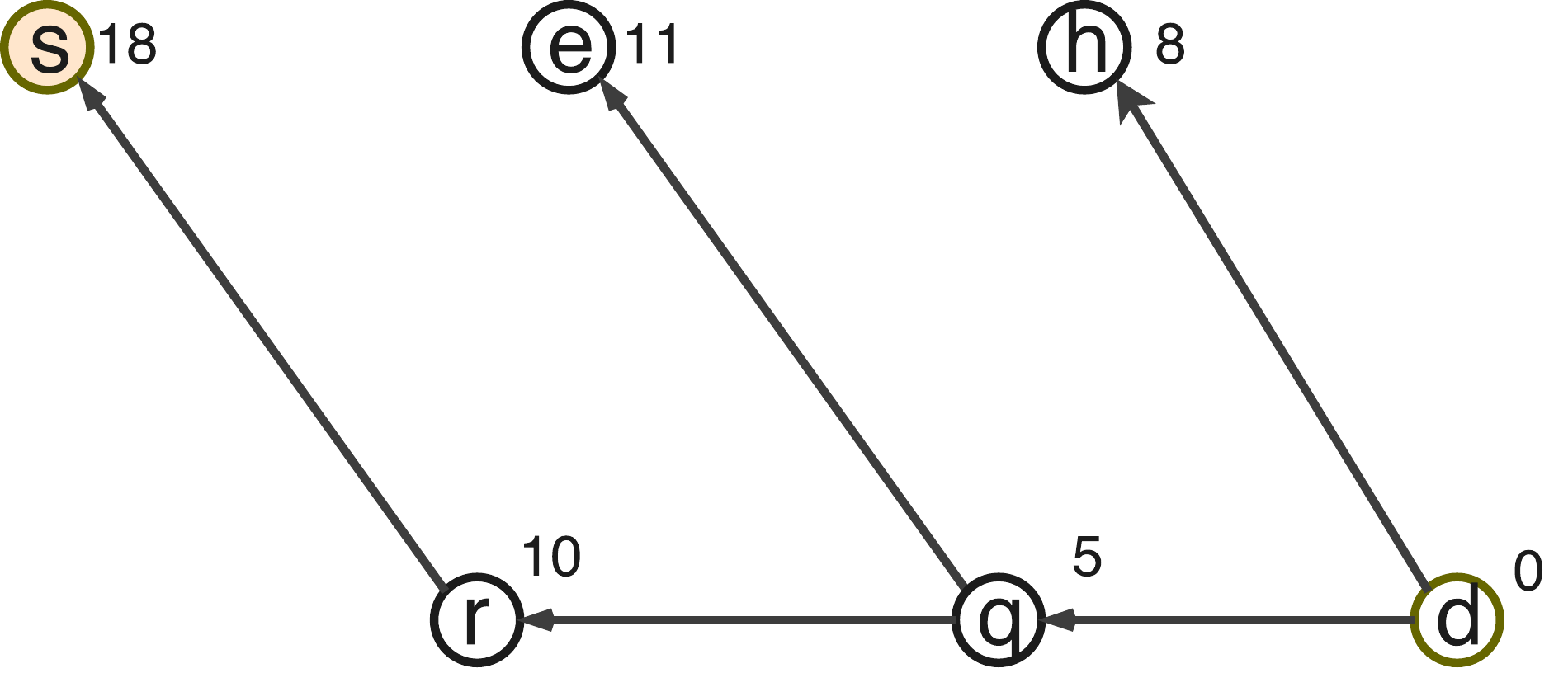}
  \end{center}
   \caption{The Reversed Shortest Path Tree With Root $d$}
  \label{fig:ui_graph}
\end{figure}

Figure~\ref{fig:ui_graph} shows the reversed shortest path tree rooted at vertex $v_d$ with budget $T=20$. If budget is 15, the shortest path tree excludes edge $\langle r, s\rangle$. And we have $v_e.getMin()=11$ meaning that it took at least 11 mins to travel from $v_e$ to destination $v_d$.

%In order to perform a Dijkstra search from destination vertex $d$ in a graph $G_{rev}$, we label all edges of the graph with deterministic values representing the minimum travel cost which is needed to travel from the start node of the edge to its end node and then we reverse the graph G. Once all arcs have been labeled with deterministic values, Dijkstra’ algorithm can be used to label all vertices which have been visited by the shortest path tree computation with their minimum travel cost to a destination node $d$.

Based on the reversed shortest path tree, we derive the ``best'' travel time distribution of reaching destination vertex $v_d$ from any vertex $v_i$ within $t$ time units as $u_i(t)$, which is shown in Equation~\ref{eqn_ui}.
\begin{equation}
\label{eqn_ui}
u_i (t)= \begin{cases}
1 &\text{if $t \geq v_i.getMin()$}\\
0 &\text{if $t<v_i.getMin()$}
\end{cases}
\end{equation}
%
%Here, $t$ is a specific travel time, $v_i \in V$
%where $v_i.GetMinVal()$ returns the minimum travel time from vertex $v_i$ to destination vertex $v_d$, i.e., the value

Given a path $P$ from source vertex $v_s$ to vertex $v_i$, we are able to compute the cost distribution of path $P$ using the PACE model.
Now, based on best travel time distribution $u_i(t)$, we are able to estimate the best travel time distribution if we follow $P$ and continue to reach the destination $v_d$.
%
%of path $P'$ that arrives at destination $v_d$ and is with $P$ as a sub-path. This means that $P'$ is from $v_s$ to $v_i$ and then to $v_d$ and $P'$ uses path $P$ from $v_s$ to $v_i$.
%
In particular, we use $r_i^{P}(x)$ to denote the largest probability of arriving the destination $v_d$ within $x$ mins if we follow path $P$ to go from $v_s$ to $v_i$ and then reach the destination $v_d$.
%
% The proposed solution calculates the travel cost distribution from a source node $s$ following the most accurate path leading to an intermediate node $i$ and consequently to a destination node $d$. At the beginning the path consists of no arcs, next incrementally, the most reliable path has been constructed step by step by adding the next arc to the path.  We denote the travel cost distribution of a path that starts in an edge $e_s$ and ends in an edge $e_d$ to be $q_{e_s,...,e_d}(t)$ in G where $t$ is a given time.
% For each intermediate vertex $i$ which belongs to a path p $<e_1,...,e_i,...,e_d>$, there must be an edge $e_i$ that ends in a node $i$. Based on this, we can calculate value representing the total travel cost distribution of a path $p=<s,$...,$i,$...,$d>$. We denote this travel cost as $r$ and defined it in Equation
% \ref{eqn_example}.
%
\begin{equation}
\label{eqn_example}
r_i^{P}(x) = \sum\limits_{k=1}^{x} prob_{P}(k).u_i(x-k)
\end{equation}
%
%Equation \ref{eqn_example} provides an estimation of the travel cost distribution from the start node $s$ following the optimal path to an intermediate vertex $i$ and then consequently to destination vertex $d$, this estimation of the travel cost has been denoted with $r$.
%
where $prob_{P}(k)$ indicates the probability that $P$ takes $k$ mins and
$u_i(x-k)$ indicates the probability that it takes less than $x-k$ mins to go from $v_i$ to $v_d$.
%
%Convolution of probability distributions has been used in the area of stochastic routing in order to calculate the sum of two independent random variables \cite{a2,a3,a4,a8}. Equation \ref{eqn_example} uses convolution of two independent random variables represented with their probability mass functions in order to derive their sum as a probability mass function.

\subsection{Proposed Algorithm}
The algorithm for computing the SPOTAR problem is shown in Algorithm~\ref{alg1}.

\begin{algorithm}
\caption{SolveSPOTAR(Vertex $v_s$, Vertex $v_d$, TimeBudget $T$)}
\label{alg1}
\begin{algorithmic}[1]
%\STATE $ \textcircled{.} $ denotes convolution and
%\STATE $ \cup $ denotes concatenation
%\\\hrulefill
%\REQUIRE $n \geq 0 \vee x \neq 0$
%\ENSURE $y = x^n$
\STATE  Initialize a priority queue $Q$; % \leftarrow PriorityQueue()$
\STATE Path $P^* \leftarrow null$; double $maxProb \leftarrow 0$;
\label{alg:line:Q_init}
\FOR{$v_i \in$ all vertices that can be reached from $v_s$}
\label{alg:line:init_for}
\STATE  $siJoint \leftarrow W(v_s,v_i)$, $siCost \leftarrow  W(v_s,v_i)$;
\label{alg:line:siJoint}
%\STATE   jointToCost(siJoint)$
\IF{$MinCost(siCost) + v_i.getMin() \leq T$}
\label{alg:line:check_out}
%\STATE $r \leftarrow  Convolution(T,siCost,u_i)$
\STATE $r \leftarrow r_i^{\langle (v_s, v_i) \rangle}(T)$ based on Eq.~\ref{eqn_example};
\label{alg:line:convolution_out}
\STATE $Q.Push(r, (\langle (v_s, v_i) \rangle, siJoint, siCost))$;
\label{alg:line:Q_push_out}
\ENDIF
\ENDFOR
\WHILE {$Q$ is not empty}
\label{alg:line:main_loop}
\STATE $(r, (P, pathJoint, pathCost)) \leftarrow Q.ExtracMax()$;
\label{alg:line:init_element}
\STATE $v_i \leftarrow$ last vertex in path $P$;
\label{alg:line:get_last_node}
\IF{$v_i == v_d$}
\label{alg:line:check_end_node}
\IF{$r>maxProb$}
	\STATE $P^* \leftarrow P$; $maxProb \leftarrow r$;
	\label{alg:line:update_best_sol}
    \STATE Delete all elements whose $r$ values are smaller than the current $r$ from $Q$;
\label{alg:line:del_all_small_sol}
\STATE \textbf{Continue;}
\label{alg:line:contin}
\ELSE
\STATE \textbf{Break;}
\ENDIF
\ENDIF
\FOR{$v_k \in$ all vertices that can be reached from $v_i$}
\label{alg:line:for_inner}
\IF{Path $P$ has not traversed $v_k$ yet}
\label{alg:line:acyclic}
\STATE $extP \leftarrow P \oplus (v_i, v_k)$;	
\STATE $ikMin \leftarrow  MinCost(v_i, v_k)$ ;\label{alg:line:ijMin}
\STATE $pathMin \leftarrow  MinCost(pathCost)$;
\label{alg:line:pathMin}
\IF{$ikMin + pathMin + j.getMin() \leq T$}
\label{alg:line:check_lower}
\STATE $extPJoint \leftarrow  PACE(extP)$;
\label{alg:line:extPJ}
\STATE $extPCost \leftarrow  JointToCost(extPJoint)$;
\label{alg:line:extPC}
\STATE $r \leftarrow  r_k^{extP}(T)$ based on Eq.~\ref{eqn_example};
\label{alg:line:conv_main_loop}
\IF{$CheckStochasticDominance()$}
\label{alg:line:CheckStochasticDominance}
\STATE $Q.Push(r, (extP, extPJoint, extPCost))$
\label{alg:line:Q_push}
\ENDIF
%\STATE $Q.Push((q_decomp$ \textcircled{.} $u_j )[T ], (qdecomp, P \cup j ))$
\ENDIF
\ENDIF
\ENDFOR
\ENDWHILE
\STATE return path $P^*$;
\label{alg:line:return_end}
\end{algorithmic}
\end{algorithm}

\begin{spacing}{1.0}
\end{spacing}

We start initializing a priority queue $Q$ that holds elements in the form of $(r, (P, pathJoint, pathCost))$, where $P$ is a path from source $v_s$ to some vertex $v_i$, $r=r_i^P(T)$ is the largest probability of arriving $v_d$ within time budget $T$ using path $P$, $pathJoint$ is the joint distribution of $P$, and $pathCost$ is the total travel time distribution of $P$.
We also initialize path $P^*$, which intends to maintain a candidate SPOTAR path and its probability $maxProb$ of arriving destination $v_d$.
%been used as explained in equation \ref{eqn_example}, $pathJoint$ is the joint distribution of a path $P$ and $pathCost$ is the cost distribution of a path $P$

We first check all vertices that are from vertex $v_s$ (line~\ref{alg:line:init_for} to line~\ref{alg:line:Q_push_out}).
%
%From  the algorithm explores all arcs that starts at node $s$,
%
Since now the candidate path $\langle (v_s, v_i) \rangle$ only has one edge, the joint distribution and cost distribution are the same (line \ref{alg:line:siJoint}). Next, if the minimum travel time from the candidate path to $v_d$ is already larger than the time budget $T$, we do not insert it into the priority queue. Otherwise, we insert it.
%
%Line \ref{alg:line:check_out}
%
%we check whether they can be pruned, if not in line \ref{alg:line:convolution_out} we compute the convolution of each arc that starts in node $s$ and we push the new path in the priority queue $Q$ in line \ref{alg:line:Q_push_out}

We proceed to the main while loop at line \ref{alg:line:main_loop}: in each iteration, we extract the element with the largest $r$ value until the priority queue $Q$ is empty.

Assume that the element with the largest $r$-value represents path $P$. We check if $P$ has already arrived destination $v_d$. If yes, we check if $P$'s $r$-value is larger than the current maximum probability $maxProb$. If yes, we update the candidate SPOTAR path $P^*$ and remove all the elements from $Q$ if their $r$-values are smaller than the current $r$. If not, we can stop the algorithm since all the remaining paths in the priority queue have $r$-values less than the current maximum probability $maxProb$. This means it makes no sense to keep exploring the remaining paths, since their probabilities of arriving $v_d$ within time budget $T$ cannot be greater than $maxProb$.

On the other hand, if $P$ has not arrived at $v_d$ yet, i.e., $v_i\neq v_d$, we need to extend $P$ by one more edge $(v_i, v_k)$ to get an extended path $extP$, which arrives vertex $v_k$.
Here, we only consider simple path without cycles as it has been shown that a path with cycles cannot be a SPOTAR path~\cite{a4}.

If it is possible to follow $extP$ to arrive $v_d$ within the time budge $T$, we compute its joint distribution $extPJoint$ using the PACE model and its total cost distribution $extPCost$. %Otherwise, we skip the extended path $extP$.
We next apply Equation~\ref{eqn_example} to compute the $r$- value of the extend path $extP$.

Before inserting the extended path $extP$ into the priority queue $Q$, we need to conduct the following stochastic dominance check.
If there already exists a path $P_k$ in $Q$ which also arrives
$v_k$. We check if $extP$'s cost distribution \emph{stochastic dominanates}~\cite{a6} $P_k$'s cost distribution. If yes, we remove $P_k$'s element from $Q$ and insert $(r$, $(extP$, $extPJoint$, $extPCost))$ into $Q$. Similarly, if $P_k$ stochastic dominates $extP$, we do not need to insert $extP$ into $Q$. Note that only considering if $extP$'s $r$-value is larger than $P_k$'s $r$-value is insufficient~\cite{a2, a6}.

Finally, we return the SPOTAR path $P^*$ along with the maximum probability of arriving $v_d$ within time budget $T$.

\section{Running example}
\label{running_example}

This section presents a concrete example of the SPOTAR problem which is solved by the proposed algorithm.
The graph from Figure \ref{fig:origin_graph} is used to show a running example of the Algorithm for computing SPOTAR under the \emph{PACE} model~\cite{a0}. The joint distribution and the derived cost distribution for paths $<e_1 , e_4>$ and $<e_2 , e_6>$ for the graph in Fig.~\ref{fig:origin_graph} are shown in Tables\ref{tbl:PACEe1e4} and \ref{tbl:PACEe2e6}, respectively. For example, the joint distribution for path $<e_1,e_4>$ has been derived from the trajectories shown in Table \ref{tbl:traj}.

\begin{spacing}{1.5}
\end{spacing}

\begin{table}[!htp]
	\centering
	\subfloat[$W(P_{<e_1,e_4>})$]{
		\begin{minipage}[c]{0.24\textwidth}
			\begin{tabular}{|c|c|c|c|} \hline
             $e_1$ & $e_4$ & Probability \\ \hline \hline
             8 & 6 & 0.8 \\ \hline
             10 & 10 & 0.2 \\ \hline
            \end{tabular}
		\end{minipage}
	}
	\subfloat[Total Cost Distribution]{
		\begin{minipage}[c]{0.24\textwidth}
			\begin{tabular}{|c|c|} \hline
             $P_{e_1,e_4}$ & Probability \\ \hline \hline
             14 & 0.8 \\ \hline
             20 & 0.2 \\ \hline
            \end{tabular}
		\end{minipage}
	}
	\caption{Joint vs. total cost distributions for path $<e_{1},e_{4}>$}
	\label{tbl:PACEe1e4}
\end{table}

\begin{table}[!htp]
	\centering
	\subfloat[$W(P_{<e_2,e_6>})$]{
		\begin{minipage}[c]{0.24\textwidth}
			\begin{tabular}{|c|c|c|c|} \hline
             $e_2$ & $e_6$ & Probability \\ \hline \hline
             8 & 5 & 0.7 \\ \hline
             11 & 9 & 0.3 \\ \hline
            \end{tabular}
		\end{minipage}
	}
	\subfloat[Total Cost Distribution]{
		\begin{minipage}[c]{0.24\textwidth}
			\begin{tabular}{|c|c|} \hline
             $P_{e_2,e_6}$ & Probability \\ \hline \hline
             13 & 0.7 \\ \hline
             20 & 0.3 \\ \hline
            \end{tabular}
		\end{minipage}
	}
	\caption{Joint vs. total cost  distributions for path $<e_{2},e_{6}>$}
	\label{tbl:PACEe2e6}
\end{table}

\textit{Problem:} Consider a problem that aims at finding the path with the highest probability of reaching destination $d$
from source $s$ in a graph $G$ within a time budget $T=22$. In addition, we have $\delta = 1$; and the following paths' joint distributions
are known a priori and is maintained in the path weight function $W$: $<e_1 , e_4>$ in Table \ref{tbl:PACEe1e4}, $<e_2 , e_6 >$ in Table \ref{tbl:PACEe2e6}

\textit{Solution:}
The reversed graph with minimum travel times for all edges is shown in Figure \ref{fig:min_travel_graph}.
%
%\begin{figure}[!htp]
%  \begin{center}
%\includegraphics[width=70mm,scale=0.5]{graph_min_travel_time.png}
%  \end{center}
%   \caption{The graph with minimum travel time on all edges}
%  \label{fig:min_travel_graph}
%\end{figure}
%
Next, the algorithm computes the shortest path tree based on the graph, and labels each node with the distance to destination node $d$ as shown in Figure $\ref{fig:ui_graph}$.
%
%\begin{figure}[!htp]
%  \begin{center}
%\includegraphics[width=67mm,scale=0.5]{graph_ui.png}
%  \end{center}
%   \caption{The graph with all $u_i$ values computed}
%  \label{fig:ui_graph}
%\end{figure}
%

For each node $i$, we define $u_i(t)$ as the highest probability of reaching the destination node $d$ from node $i$, i.e., Equation~\ref{eqn_ui} as stated in \cite{a0}, where $i.GetMinVal()$ denotes the minimum travel time from node $i$ to destination node $d$ and $t$ is a given time budget.

\begin{equation}
\label{eqn_ui}
u_i (t)= \begin{cases}
1 &\text{if $t \geq v_i.getMin()$}\\
0 &\text{if $t<v_i.getMin()$}
\end{cases}
\end{equation}

The algorithm begins with initialization of the priority queue $Q$. Next, it iterates over all adjacent arcs of $s$, i.e. $e_1$ and $e_2$. For arc $e_1$ the algorithm computes $prob_{e_1}$ which is given as follows:
\begin{spacing}{1.5}
\end{spacing}

$ prob_{e_1} = $\{$(8,0.9),(10,0.1)$\}

\begin{spacing}{1.5}
\end{spacing}

Next, the value of $r$ is computed using $u_e$ as follows:

\begin{spacing}{1.5}
\end{spacing}

$r = \sum\limits_{k=1}^{T} prob_{ei}(k ).u_e(T-k) = \sum\limits_{k=1}^{T=22} prob_{e_1}(k ).u_{e}(22-k) =  prob_{e_1}(8).u_e(22-8) + prob_{e1}(10).u_e(22-10)=0.9*1+0.1*1=1$

\begin{spacing}{1.5}
\end{spacing}

After the value of $r$ has been computed, edge $e_1$ has been added to the path created until now, i.e., $<e_1>$. The path along with its $r$ value are pushed to the priority queue $Q$.

\begin{spacing}{1.5}
\end{spacing}
$Q.push( 1.0, (<e_1>,prob_{e_1},  cost_{e_1}))$
\begin{spacing}{1.5}
\end{spacing}

The second adjacent arc of node $s$ that has been considered by the algorithm is arc $e_2$. The algorithm again starts by computing the joint probability for the path discovered until now (i.e. $e_2$), which is shown as follows. % It has been given as follows:
\begin{spacing}{1.5}
\end{spacing}

$ prob_{e_2} = $\{$(8,0.2),(11,0.8)$\}

\begin{spacing}{1.5}
\end{spacing}

Next, the algorithm computes $r$ values for node $r$ using $u_r = 10$.

\begin{spacing}{1.5}
\end{spacing}
$r = \sum\limits_{k=1}^{T} prob_{e_2}(k ).u_r(T-k) = \sum\limits_{k=1}^{T=22} prob_{e_2}(k ).u_{e}(22-k) =  prob_{e_2}(8).u_r(22-8) + prob_{e_2}(11).u_r(22-11)=0.2*1+0.8*1=1$
\begin{spacing}{1.5}
\end{spacing}

Once the r-value of $r$ has been computed, edge $e_2$ has been added to the path created until now and its value has been pushed to the priority queue $Q$
\begin{spacing}{1.5}
\end{spacing}
$Q.push( 1.0, (<e_2>,prob_{e_2},cost_{e_2}))$
\begin{spacing}{1.5}
\end{spacing}

Next, the algorithm has to pop the value with the maximum key from the priority queue $Q$. Until now there are two elements in $Q$ i.e. edge $e_1$ and edge $e_2$ both with key value equal to 1. The algorithm decides to pop the element which is associated with the path $e_1$ based on the key $(1.0) $.

After that, the algorithm iterates over all adjacent arcs of node $e$ i.e. arcs $e_5$ and $e_4$.

For arc $e_5$ conditional independence holds because there are no overlapping sub-paths that contains $e_5$ and the path constructed so far by the algorithm. Therefore, the algorithm computes the joint distribution and then derive the associated cost distribution as shown in Table \ref{tbl:PACEe1e5}.

$prob_{e_1,e_5} = prob_{e_1} \cdot prob_{e_5}$

\begin{table}[!htp]
	\centering
	\subfloat[$W(P_{e_1,e_5})$]{
		\begin{minipage}[c]{0.24\textwidth}
			\begin{tabular}{|c|c|c|c|} \hline
             $e_1$ & $e_5$ & Probability \\ \hline \hline
             8 & 8 & 0.72 \\ \hline
             8 & 10 & 0.18 \\ \hline
             10 & 8 & 0.08 \\ \hline
             10 & 10 & 0.02 \\ \hline

            \end{tabular}
		\end{minipage}
	}
	\subfloat[Total Cost Distribution]{
		\begin{minipage}[c]{0.24\textwidth}
			\begin{tabular}{|c|c|} \hline
             $P_{e_1,e_5}$ & Probability \\ \hline \hline
             16 & 0.72 \\ \hline
             18 & 0.26 \\ \hline
             20 & 0.02 \\ \hline

            \end{tabular}
		\end{minipage}
	}
	\caption{Joint and the derived cost  distributions for path $<e_{1},e_{5}>$}
	\label{tbl:PACEe1e5}
\end{table}

\begin{spacing}{1.5}
\end{spacing}

Next, the algorithm prunes path $<e_1,e_5>$ based on minimum travel time to destination $d$ i.e. $(16+8) > T = 22$, therefore nothing has been added to priority queue.

The next adjacent arc of the path  $<e_1>$ that ends in node $e$ is $e_4$. It passes through path $<e_1,e_4>$, therefore the algorithm uses the given joint and cost distributions from Table \ref{tbl:PACEe1e4} and then $r$ has been computed, where $u_q = 5$.

\begin{spacing}{1.5}
\end{spacing}

$r = \sum\limits_{k=1}^{T} prob_{e_1,e_4}(k ).u_q(T-k) = \sum\limits_{k=1}^{22} prob_{e_1,e_4}(k ).u_{q}(22-k) =  prob_{e_1,e_4}(14).u_q(22-14) + prob_{e_1,e_4}(20).u_q(22-20)= 0.8*1+0.2*0=0.80$

\begin{spacing}{1.5}
\end{spacing}

The path $<e_1,e_4>$ has been added to the priority queue $Q$.
\begin{spacing}{1.5}
\end{spacing}

$Q.push( 0.8, (<e_1,e_4>,prob_{e1,e4},  cost_{e1,e4}))$

\begin{spacing}{1.5}
\end{spacing}

Until now, there are two paths in the priority queue i.e. path $<e_2>$ and path $<e_1,e_4>$ with keys 1.0 and 0.8, respectively.

The algorithm pops the element associated with path $<e_2>$ from the priority queue based on the element keys since $1.0 > 0.8$.

Then the algorithm iterates over all adjacent arcs of $r$ i.e. $e_3$ and $e_6$. Arc $e_3$ can be pruned based on minimum travel time calculated as follows $(8+11) + 11 > T = 22$. Since arc $e_6$ passes through path $<e_2,e_6>$, the algorithm uses the given probability in Table \ref{tbl:PACEe2e6} in order to compute a value of $r$, where $u_q = 5$ as follows:

\begin{spacing}{1.5}
\end{spacing}

$r = \sum\limits_{k=1}^{T} q_{e_2,e_6}(k ).u_q(T-k) = \sum\limits_{k=1}^{22} prob_{e_2,e_6}(k ).u_{q}(22-k) =  prob_{e_2,e_6}(13).u_q(22-13) + prob_{e_2,e_6}(20).u_q(22-20)= 0.7*1+ 0.3*0 = 0.70 $

\begin{spacing}{1.5}
\end{spacing}

Since path $<e_2,e_6>$ ends in node $q$ and path $<e_1,e_4>$ also ends in node q and it is already added to the priority queue, we try to prune one of the paths based on Stochastic dominance as explained in \cite{a0}.

\begin{figure}[!htp]
  \begin{center}
\includegraphics[width=85mm]{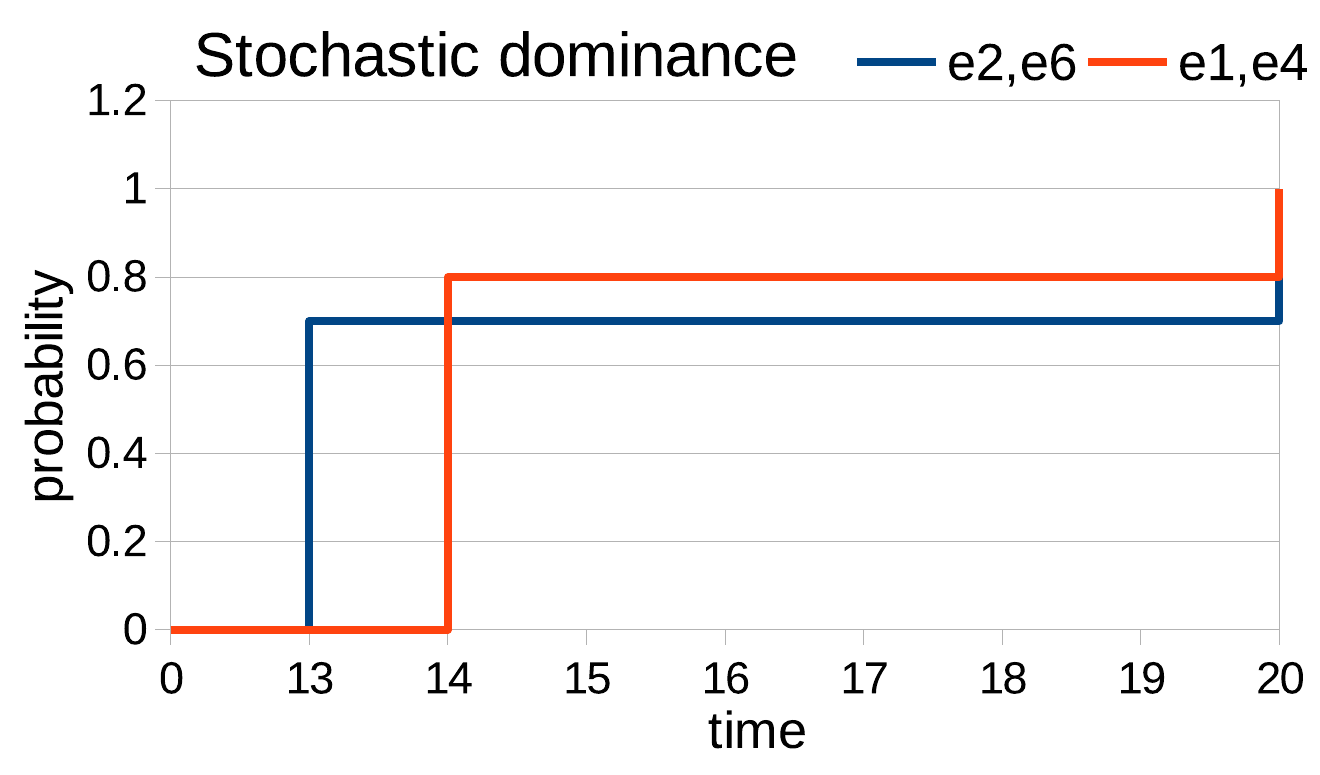}
  \end{center}
   \caption{Stochastic dominance between paths $<e_1,e_2>$ and $<e_2,e_6>$. None of the paths stochastically dominates the other one.}
\label{fig:stochastic_dominans}
\end{figure}

As it can be seen from Figure \ref{fig:stochastic_dominans}, none of the path stochastically dominates the other, therefore we cannot prune either of them, hence the algorithm adds path $<e_2,e_6>$  to the priority queue $Q$

\begin{spacing}{1.5}
\end{spacing}

$Q.push( 0.70, ( <e_2 ,e_6>,prob_{e_2,e_6}, cost_{e_2,e_6}))$

\begin{spacing}{1.5}
\end{spacing}

At this point the priority queue $Q$ contains two paths: $<e_1,e_4>$  and $<e_2,e_6>$  with the following objects associated as follows:

\begin{spacing}{1.5}
\end{spacing}

$( 0.80, ( <e_1, e_4 >,prob_{e_1,e_4} , cost_{e_1,e_4} )$

$( 0.70, ( <e_2 ,e_6 >,prob_{e_2,e_6} ,cost_{e_2,e_6})$

\begin{spacing}{1.5}
\end{spacing}

Path $<e_1,e_4>$  has been popped up from the priority queue $Q$ based on its key 0.8 and for all adjacent edges of node $q$ i.e. arcs $e_7$ and $e_9$.

Arc $e_7$ is pruned by the algorithm bases on its minimum travel time to destination calculated as follows: $(14+13) + 8 > T = 22$

Arc $e_9$ is explored by the algorithm, it computes the travel cost distribution for path $<e_1 , e_4 , e_9 >$ where independence between $<e_1 , e_4>$ and $<e_9>$ does hold, the result can be seen in Table \ref{tbl:PACEe1e4e9}

\begin{table}[!htp]
	\centering
	\subfloat[$W(P_{e_1,e_4,e_9})$]{
		\begin{minipage}[c]{0.24\textwidth}
			\begin{tabular}{|c|c|c|c|} \hline
             $e_1$ & $e_4$ & $e_9$ & Probability \\ \hline \hline
             8 & 6 & 5 & 0.32 \\ \hline
             8 & 6 & 9 & 0.48 \\ \hline
             10 & 10 & 5 & 0.08 \\ \hline
             10 & 10 & 9 & 0.12 \\ \hline
            \end{tabular}
		\end{minipage}
	}
	\subfloat[Total Cost Distribution]{
		\begin{minipage}[c]{0.24\textwidth}
			\begin{tabular}{|c|c|} \hline
             $P_{e_1,e_4,e_9}$ & Probability \\ \hline \hline
             19 & 0.32 \\ \hline
             23 & 0.48 \\ \hline
             25 & 0.08 \\ \hline
             29 & 0.12 \\ \hline
            \end{tabular}
		\end{minipage}
	}
	\caption{Joint vs. total cost distributions for path $<e_1,e_4,e_9>$}
	\label{tbl:PACEe1e4e9}
\end{table}

The search reaches a destination node $d$. Solution 1 for path $< e_1 , e_4 , e_9 >$ with probability 0.32 of reaching the destination $d$ within a time budget of 22 units.
After obtaining a solution. The algorithm checks the keys (i.e., the r-values) of all elements in priority queue. If the key is less then the probability of the solution $(0.32)$, we prune the corresponding entry from the priority queue. Since $0.70>0.32$, we therefore cannot prune path $<e_2, e_6>$  which is in the priority queue.

Pop $<e_2,e_6>$ based on key $0.70$. For all adjacent arcs of node $q$
First $e_7$ can be pruned based on minimum travel time $(13+13) + 8 > T = 22$
Second, arc $e_9$. The algorithm computes the travel cost for the path $<e_2 , e_6 , e_9 >$. Independence does hold, therefore therefore we have that $prob_{e_2,e_6,e_9} = prob_{e_2,e_6} \cdot prob_{e_9}$ the result is shown in Table \ref{tbl:PACEe2e6e9}

\begin{table}[!htp]
	\centering
	\subfloat[$W(P_{e_2,e_6,e_9})$]{
		\begin{minipage}[c]{0.24\textwidth}
			\begin{tabular}{|c|c|c|c|} \hline
             $e_2$ & $e_6$ & $e_9$ & Probability \\ \hline \hline
             8 & 5 & 5 & 0.28 \\ \hline
             8 & 5 & 9 & 0.42 \\ \hline
             11 & 9 & 5 & 0.12 \\ \hline
             11 & 9 & 9 & 0.18 \\ \hline
            \end{tabular}
		\end{minipage}
	}
	\subfloat[Total Cost Distribution]{
		\begin{minipage}[c]{0.24\textwidth}
			\begin{tabular}{|c|c|} \hline
             $P_{e_2,e_6,e_9}$ & Probability \\ \hline \hline
             18 & 0.28 \\ \hline
             22 & 0.42 \\ \hline
             25 & 0.12 \\ \hline
             29 & 0.18 \\ \hline
            \end{tabular}
		\end{minipage}
	}
	\caption{Joint vs. total cost  distributions for path $<e_2,e_6,e_9>$}
	\label{tbl:PACEe2e6e9}
\end{table}

The search reaches destination node d. Solution $2$ for path $< e_2 ,e_6 ,e_9 >$ with probability $0.70$ of reaching the destination node $d$ within a time budget $22$ units. We check the keys of all elements in priority queue. If the key is less then $0.70$, we eliminate the corresponding entry from the priority queue. Priority queue is empty, so we can not prune.
Since solution 2 with path $< e_2 ,e_6 ,e_9 >$ has higher probability of reaching the destination node $d$ within a time budget of $22$ units. It is the final solution.
\begin{spacing}{1}
\end{spacing}

\section{Experimental results}

We report on experimental results using real world GPS trajectories.

\subsection{Experimental Setup}
\begin{figure}[!htp]
  \begin{center}
\includegraphics[width=90mm,scale=0.4]{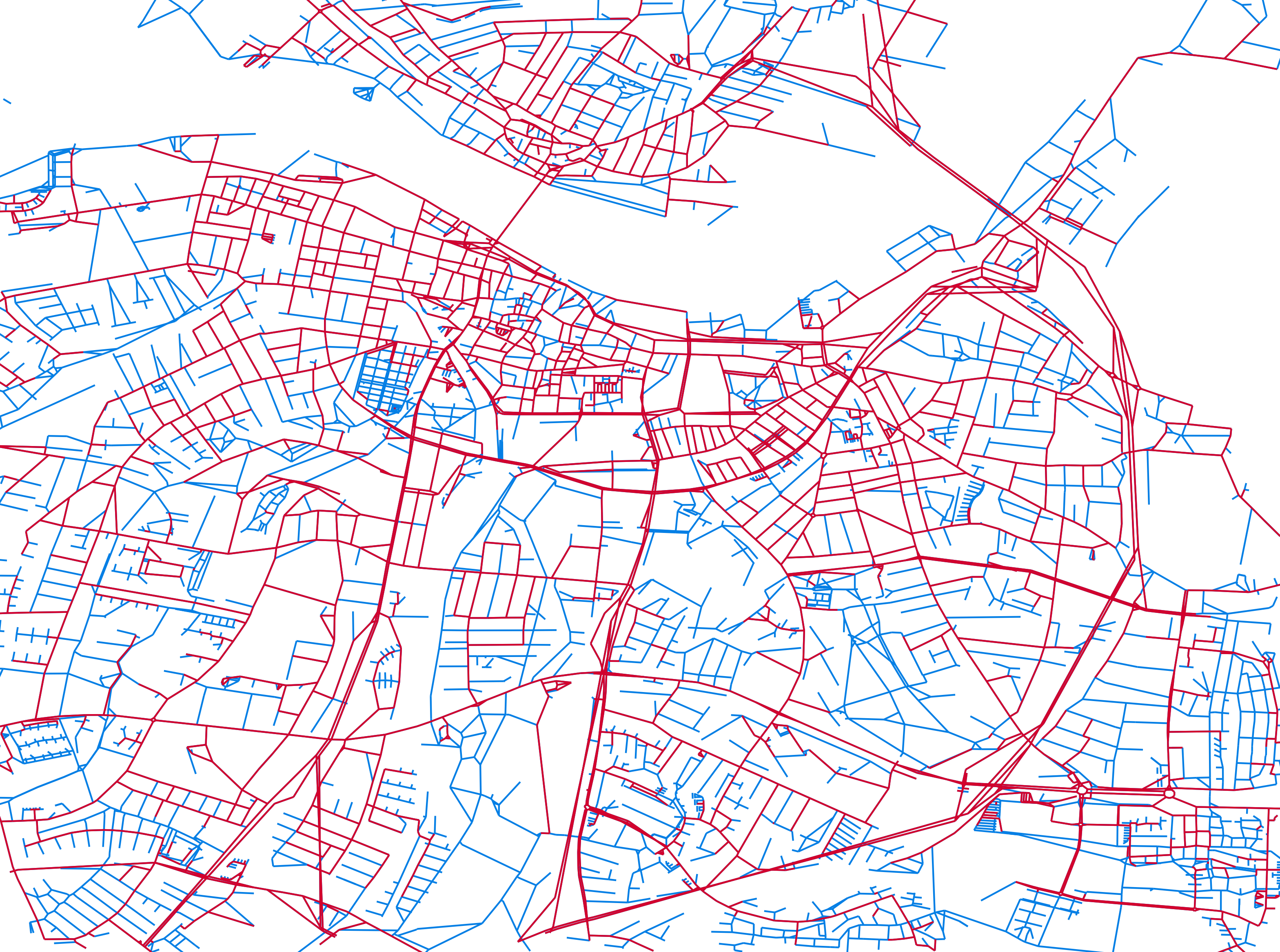}
  \end{center}
   \caption{Aalborg network. The red segments are covered by the GPS trajectories and thus their travel cost distributions are derived from the GPS trajectories. The blue segments are not covered by the GPS trajectories and thus speed limits are used to derive travel costs.}
    \label{fig:aalborg_trips}
\end{figure}

We use the road network of Aalborg, Denmark, which contains 6,253 nodes and 10,716 edges. We use 37 million GPS records that occurred in Aalborg from Jan 2007 to Dec 2008 to instantiate the PACE model. The sampling rate of GPS records is 1 Hz (i.e., one GPS record per second).
%
% has been used which is a partition of a road network consisting of whole Denmark. We extract a partition of Aalborg using four coordinates consisting of latitude and longitude. These coordinates form,  a rectangle around Aalborg which allows us to extract all nodes and edges which are inside the rectangle.
%Visual representation of the partition of Aalborg can be seen in Figure \ref{fig:aalborg_trips}.
%
%The partition of Aalborg. The travel cost distribution has been obtained by real GPS tracking observations. The GPS data which has been used in the experiments contains.
%
% \colorbox{blue!30}{37 million GPS records that occurred in Aalborg from January}
% \colorbox{blue!30}{2007 to December 2008. The sampling rate is 1 Hz (i.e., }
% \colorbox{blue!30}{one record
% per second).}.
%
If an edge is not covered by GPS data, we use the length of the edge and the speed limit on the edge to derive a travel time. If a path has been traversed by 10 trajectories, we maintain a joint distribution for the path.
A visual representation of the edges that are covered and not covered by the GPS trajectories can be found in Figure~\ref{fig:aalborg_trips}.

\begin{spacing}{1.5}
\end{spacing}

%Figure \ref{fig:aalborg_trips} provides a visual representation of the roads that have been covered by the GPS data. The red color corresponds to edges that have been covered by the trajectory data, the blue road segments are not covered by any GPS trajectory from the dataset and therefore speed limits have been used to derived travel cost.
%Figure \ref{fig:aalborg_trips} provides evidence that all major roads as well as some percentage of the rest of the roads are covered by the GPS trajectory data, however some roads are not covered by the data.

\textbf{\textit{Queries:}} We consider different settings to generate SPOTAR queries.
%
%
%
%For each of the experiments, 20 source-destination pairs were randomly picked. All 20 pears have been chosen with a positive probability of arriving on time from source to destination.
%
%\textbf{\textit{Time budget:}}
We vary time budget (seconds) from 300, 500, 700, to 1,000.
We also vary the Euclidean distance (km) between source-destination pairs: $[0, 1)$,
$[1, 2)$,  $[2, 3)$, and $[3, 4)$.
%
%More precisely we used 300, 500, 700, 1000 seconds as a value for the time budget during the evaluation of the algorithm.
%
For each setting, we randomly generate 20 source-destination pairs.

\textbf{\textit{Methods:}}
We consider different heuristic functions: (1) the proposed solution with minimum travel time to the destination using shortest path trees (SP); (2) the baseline heuristic using Euclidean distance divided by the maximum speed limit (BA).
We apply the both heuristics on both PACE and the edge-centric model (EDGE), resulting four methods: SP+PACE, SP+EDGE, BA+PACE, and BA+EDGE.
%Our proposal uses the minimum travel cost to destination as a lower bound heuristic as it was explained in \ref{su:heuristic} on p. \pageref{su:heuristic}. With intention to compare our heuristic approach we evaluate our proposal using  euclidean distance between two nodes, divided by the maximum speed of the network as  we refer this heuristic as Euclidean heuristic. We evaluate our proposal using combination of the two heuristics to gain valuable inside of the algorithm behavior.

\textbf{\textit{Evaluation Metrics:}}
We report average runtimes and sizes of search space for running SPOTAR in different settings.

We conduct experiments on a computer with Intel® Core™ i5-4210U CPU @ 1.70GHz × 4 processors with 12GB RAM under 64 bit Linux Fedora 25.
%
%The experiments have been performed on Intel® Core™ i5-4210U CPU @ 1.70GHz × 4 processors with 12GiB of Random-access memory (RAM). The operation system that has been used in the experiment is a 64 bit Linux Fedora 25.
%
The code was implemented in Python 3.

% \begin{spacing}{1.5}
% \end{spacing}

% \begin{figure}[!htp]
%   \begin{center}
% \includegraphics[width=90mm,scale=0.4]{aalborg_trips.png}
%   \end{center}
%    \caption{Aalborg network. The red segments have been rendered thanks to the trajectory GPS data. For these segments exists enough information to derived travel cost distribution. For the blue segments the speed limits have been used to derive travel cost.}
%     \label{fig:aalborg_trips}
% \end{figure}

\subsection{Experimental Results}

%Experiments have been evaluating the performance of the proposed algorithm. Combinations of different time budgets and heuristics have been used with the proposed algorithm in order to gain valuable information how the proposal has been affected and also to identify possible strengths and weaknesses.

%For each of the experiments, 20 source-destination pairs were randomly picked. All 20 pears have been chosen with a positive probability of arriving on time from source to destination.

\begin{figure}[!htp]
\centering
\subfloat[Time budget 400]{%
\includegraphics[width=0.23\textwidth]{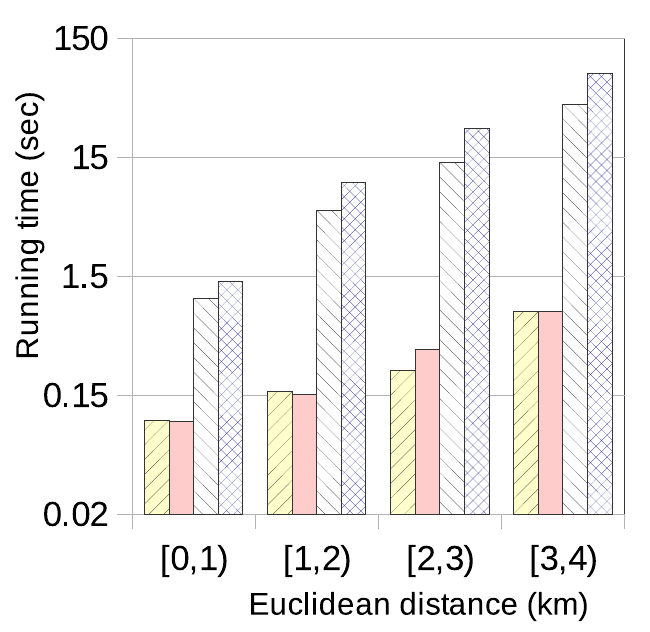}%
}%
\subfloat[Time budget 600]{%
\includegraphics[width=0.23\textwidth]{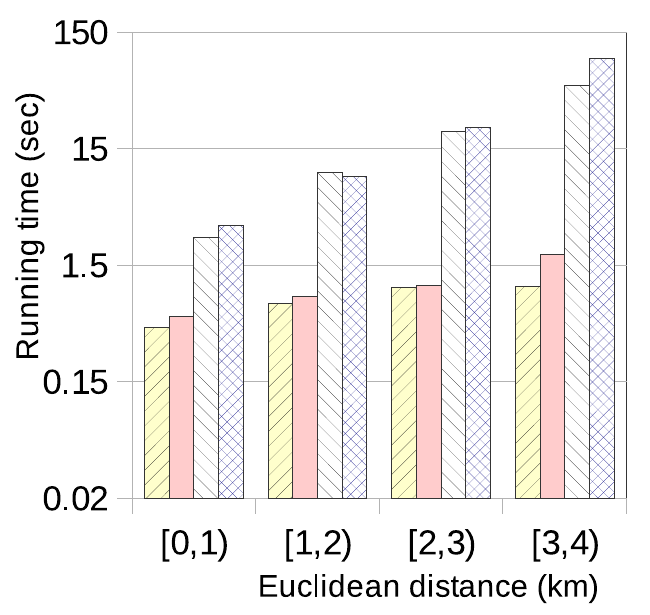}%
}%

\subfloat[Time budget 800]{%
\includegraphics[width=0.23\textwidth]{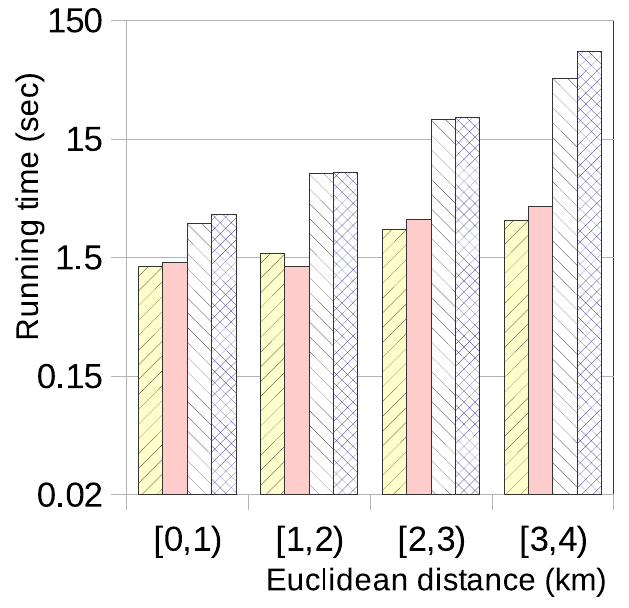}%
}%
\subfloat[Time budget 1000]{%
\includegraphics[width=0.23\textwidth]{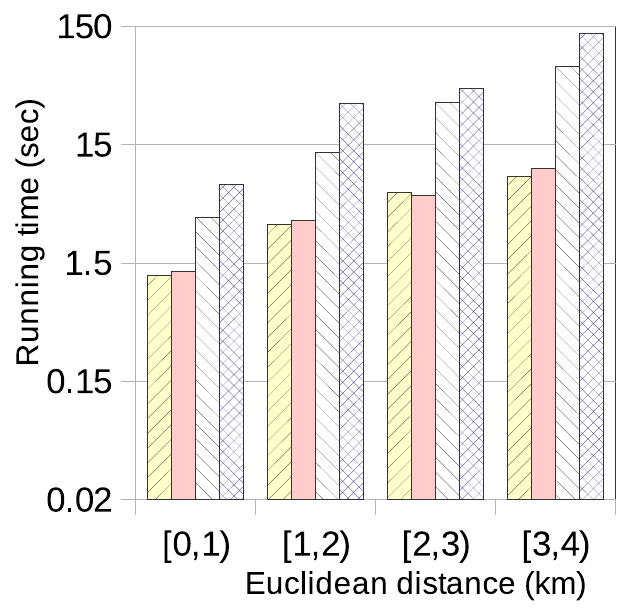}%
}%
\hfill%
\includegraphics[width=0.40\textwidth]{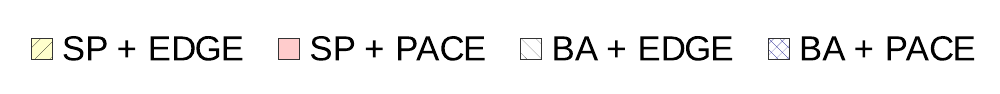}%
%\caption{Running time depending on the time budget and also considering two heuristic approaches i.e. minimum travel cost to destination and euclidean distance divided by the maximum speed in the network}
\caption{Runtime}
\label{fig:exp_heuristics}
\end{figure}

%\subsection*{Runtime}
\noindent
\textbf{Runtime: }
%We evaluate the running time of the algorithm using 4 time budgets i.e. 300,500,700,1000 seconds and also 4 euclidean distance ranges i.e. 0-1, 1-2, 2-3, 3-4 kilometers.
%
%Figure \ref{fig:exp_1} shows runtimes of the proposed algorithm.
%
%When the distance between a source-destination pair increases, the runtime of the algorithm also increases for all time budgets.
%
% Note that when time budget is 500, it  seconds solutions for distances in the range 3 to 4 kilometers can not be obtain. Similarly, for time budget equal to 300 seconds solutions in the range 2 to 3 and 3 to 4 kilometers can not be found.
% \begin{figure}[!htp]
%   \begin{center}
% \includegraphics[width=80mm]{exp_1.png}
%   \end{center}
%    \caption{Evaluating the running time of the algorithm for solving SPOTAR problem considering four time budgets and also euclidean distance between source and destination.}
%     \label{fig:exp_1}
% \end{figure}
%
%The running time of the algorithm has been also evaluated against two heuristics i.e. minimum travel cost and euclidean heuristic. The results have been presented in
Figure \ref{fig:exp_heuristics} shows the runtimes when using four methods under different settings.
When the distance between a source-destination pair increases, the runtimes of all methods also increases for all time budgets.
The SP heuristic shows significantly better runtime in comparison with the BA heuristic under all settings on both models. %This is is true for all time budgets.
In addition, it is clear than the runtime growth of the BA heuristic is much faster in comparison with the SP heuristic as the distance between a source and a destination increases.
%
%This suggests that the SP heuristic is
%
Under the same heuristic function and time budget, methods based on PACE have similar runtime with the methods based on EDGE. This suggests that although PACE is able to provide more accurate results that EDGE, it does not takes longer run time. This suggests that PACE is both accurate and efficient.
%Conclusion can be made that the selection of heuristic can be a crucial consideration in stochastic routing. %concern with the running time of the algorithm.

\noindent
%\subsection*{Search space}
\textbf{Search space: }We investigate the sizes of search spaces that different methods explore.
In particular, we define the search space as the edges that have been explored by a method.

\begin{figure}[!htp]
\centering
\subfloat[Time budget 400]{%
\includegraphics[width=0.23\textwidth]{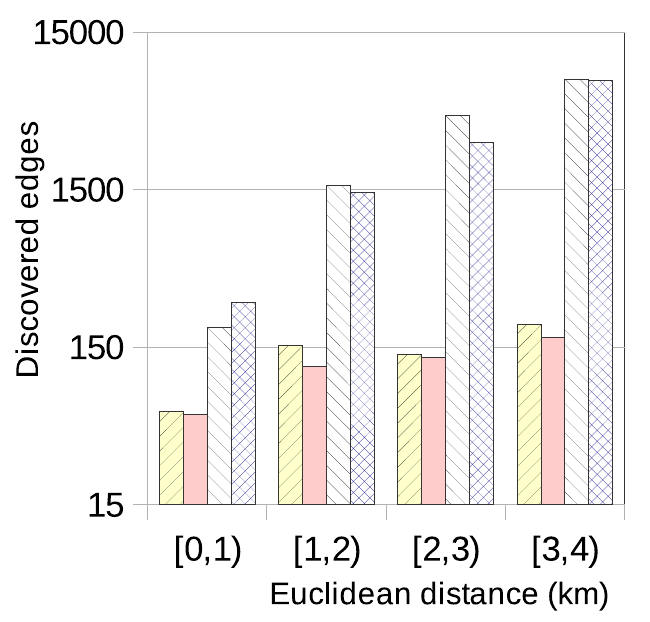}%
}%
\subfloat[Time budget 600]{%
\includegraphics[width=0.23\textwidth]{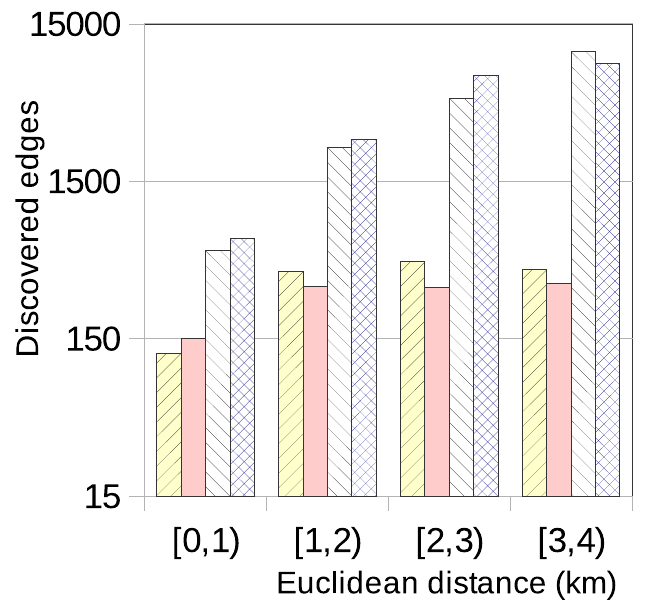}%
}%

\subfloat[Time budget 800]{%
\includegraphics[width=0.23\textwidth]{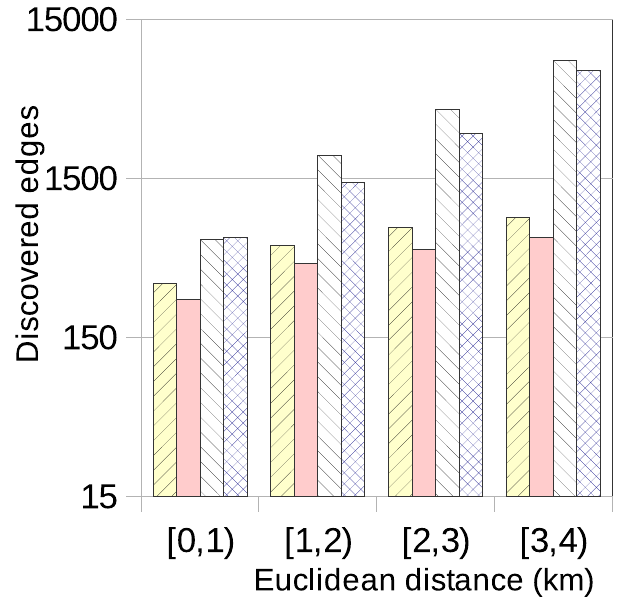}%
}%
\subfloat[Time budget 1000]{%
\includegraphics[width=0.23\textwidth]{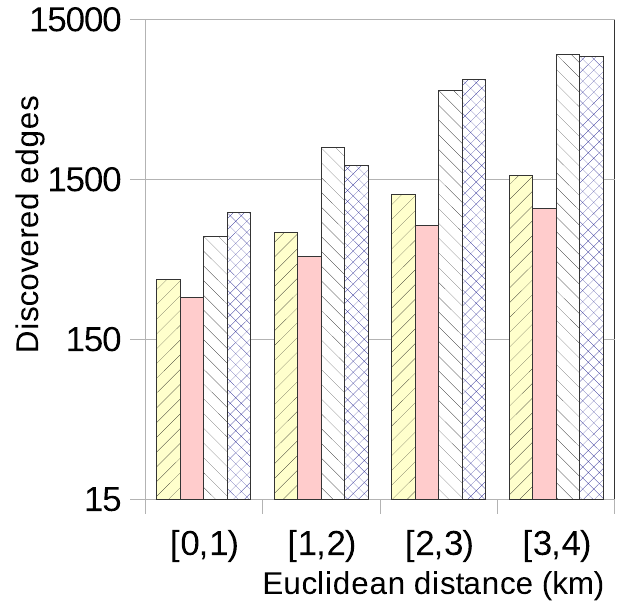}%
}%
\hfill%
\includegraphics[width=0.40\textwidth]{running_time_legend.pdf}%
%\caption{Number of edges discovered depending on the time budget and considering combinations between two heuristic approaches i.e. SP and BA, and the PACE model as well as the edge centric model(EDGECE).}
\caption{Search Space}
\label{fig:exp_heuristic_non_trivial}
\end{figure}

We compare the search spaces that are explored by different methods. % algorithms using both heuristics.
Figure~\ref{fig:exp_heuristic_non_trivial} shows that the search space of BA is much higher than that of SP in all settings, which is consistent with the performance of runtime shown in Figure~\ref{fig:exp_heuristics}.

\begin{figure}[!htp]
\centering
\subfloat[SP+PACE]{%
\includegraphics[width=0.23\textwidth]{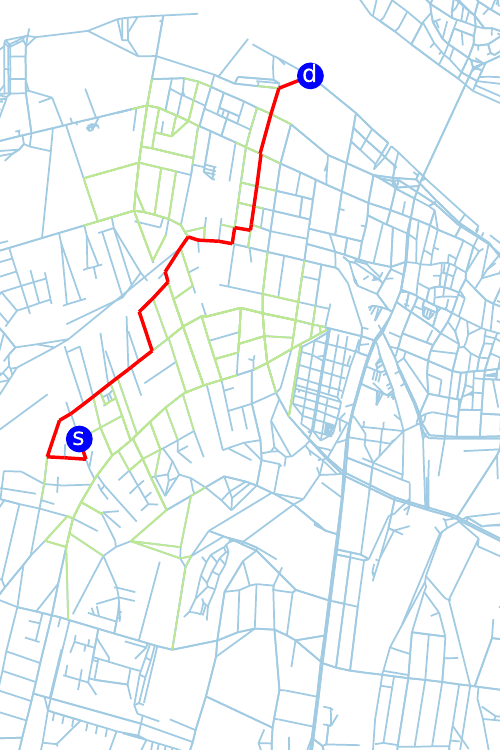}%
}%
\hfill%
\subfloat[BA+PACE]{%
\includegraphics[width=0.23\textwidth]{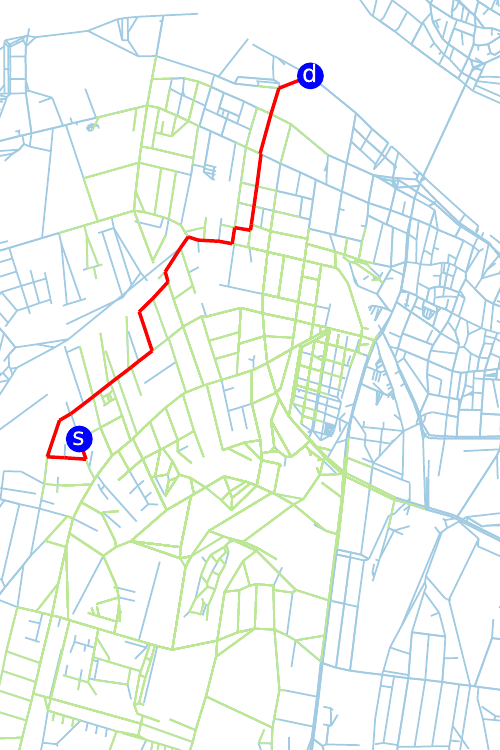}%
}%

\subfloat[SP+EDGE]{%
\includegraphics[width=0.23\textwidth]{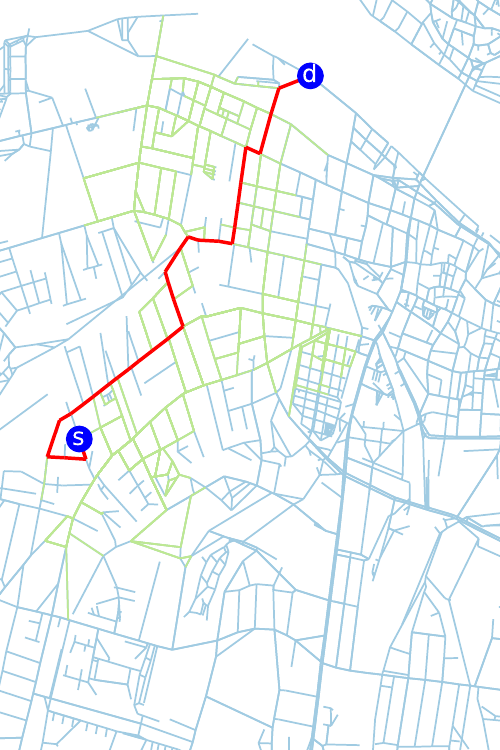}%
}%
\hfill%
\subfloat[BA+EDGE]{%
\includegraphics[width=0.23\textwidth]{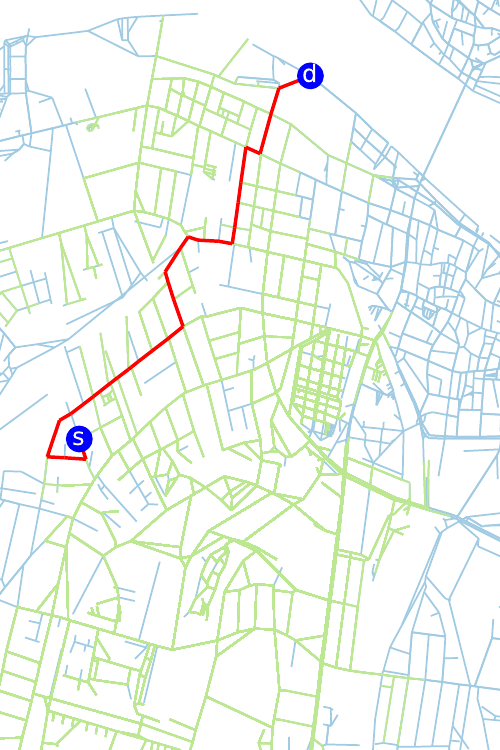}%
}%
%\caption{Pruning of Aalborg network, considering random source-destination pair in the range [3,4), time budget of 800s. Combinations between the two heuristics i.e. SP, BA, and the  models PACE, edge-centric(EDGECE).}
\caption{A specific $(s, d)$ pair, distance [3, 4), time budget 800}
\label{fig:pruning}
\end{figure}

%During the evaluation of the algorithm we have been keeping the amount of edges which the algorithm explores. The results of the evaluation using minimum travel cost as heuristic have been presented in Figure \ref{fig:exp_discovered_edges_minimum}. The results are classified into four euclidean distance ranges as well as four different time budgets. The highest amount of edges have been discovered for time budget equal to 1000 seconds, this is true for all distances. Solutions for time budget equal to 300 and 500 seconds in the range 3 to 4 kilometers can not be obtained. This also holds for time budget of 300 seconds and range 2 to 3 kilometers.

%The total amount of edges discovered by the algorithm using minimum travel cost as heuristic, classified by four time budgets can be seen in Figure \ref{fig:exp_discovered_edges}. The results from the figure suggest that the grow of the number of edges which have been discovered is exponential when the time budget has been increased.

Next, we show visually the edges that are discovered by both heuristic under a specific source-destination pair setting (see the caption of Fig.~\ref{fig:pruning}). It is clear that: (1) when using the same model, SP explores much less edges than BA does, indicating the SP heuristic is effective (see the green edges in Fig.~\ref{fig:pruning}); (2) the path returned by the PACE model is different from the path returned by the EDGE model (see the red paths in Fig.~\ref{fig:pruning}), indicating that the more accurate path distributions captured by PACE do make a difference on the returned paths.

\section{Conclusions and Outlook}
Arriving on time is an important problem that has many applications in intelligent transportation systems. %factor which has been linked to many real world problem domains.
We present an effective algorithm that solves the SPOTAR problem on the novel path-centric PACE model that considers the dependencies among travel times on different edges. %between segments in a path.
Experimental results on real-world trajectories suggest that the proposed algorithm is effective. %provide useful information about the performance of the proposed algorithm.
In the future, we plan to study further speed-up strategies, e.g., using contraction hierarchies and hub labeling, on top of the PACE model, also possibly using parallel computing~\cite{a9}.

%The presented article provides solutions within some bounds considering the parameters that have been used. Future work could consider the usage of larger time budgets, new heuristics and possible upper bounds for the search.

% \begin{spacing}{2.0}
% \end{spacing}

%\small

\end{document}